# Influence of the variations of potential on autolocalization of the surface electrons over helium in the conditions of quasi - one-dimensionality.

A.V. Smorodin and V.A. Nikolaenko

Institute for Low Temperature Physics and Engineering,
47 Lenin Avenue, 61103 Kharkov, Ukraine

smorodin@ilt.kharkov.ua

**Abstract**

The feature of interaction electrons with neutral matter attracts large attention of physicists many directions. The exchange interaction in particular leads to creating of bubbles - spherical cavities with electron in center. Good object for such study is system of surface electrons formed over surface of dielectric with small permittivity. Investigation of the autolocalization of SE on liquid helium surface attracts high attention, because system is the extremely clear and easily governed. Bubbles have small mobility relatively of free electrons and that is used for monitoring of the transition process. As has recently been investigated the quasi-one-dimensional electrons over liquid helium form the autolocalized states at smaller gas density in compared to 3D and 2D situations. The process significantly depends from a substrate and its quality.

The autolocalization of one-dimensional electrons in a gas phase over liquid helium covering a substrate at both situations with or without the potential variations is theme of present study. The features of conductivity in presence of potential variations the authors explain by both the dynamic processes of changing the subsystem energy at appear or disappear electron autolocalization and by thermo-insulation of electron subsystem from thermostat.

**Introduction**

The interaction of the electrons with neutral matter isn't trivial and attracts large attention of physicists of many science directions, such as, condensed matter physics; nano-physics; physics of plasma and high energy. Are known next consequences of exchange interaction: the appearance of bubbles (spherical cavities with electron in center); the distortion of the matter energy spectrum; the stimulation of positron annihilation. Appearance and disappearance of the autolocalized electrons is observed in [1-3]. The main intrigue this phenomena is influence of substrates and fields. The investigations of the electron bubbles in the bulk helium and gas phase described in [1-2, 4]. Good object of same investigation is surface electrons (SE). SE is formed over smooth surface of dielectric with small permittivity [see for ex. 5-6]. The investigation of the autolocalization SE on liquid helium surface attracts high attention, because system is the extremely clear and easily governed. The electron bubble has expressed energy spectrum and can be used as bit of the quantum computer, like SE was proposed for this target in [7, 8]. It was shown that the autolocalization of the electrons take place at smaller gas density in compared to 3D and 2D situation [2, 9]. As has recently been investigated the quasi-one-dimensional (Q1D) electrons over liquid helium form autolocalization state at yet smaller gas density [10]. This process significantly depends on presence of a substrate and its quality. Bubbles have small mobility relatively of free electrons and that is used for monitoring of the transition process.

Autolocalization of one-dimensional electrons in a gas phase over liquid helium covering a substrate at both situations, with or without the variations of potential, and it's the theme of this work

## Experimental procedure

The method of low frequency electron transport for study of autolocalization of quasi-one-dimensional electrons over liquid helium was applied. The Sommer-Tanner technique based on capacitive coupling of electron subsystem with measurement electrodes is used [11]. The measurements by lock-in-amplifier in current regime on frequency 20 kHz with signal amplitude 5-30 mV$_{rms}$ are performed. The SE conductivity has been calculated on measured components of the cell signal like in [12]. The cell design and experimental steps were same as in work [10]. The cell (fig.1) includes two measurement electrodes, size each electrode is 6x12 mm. Expanding of measurement range and additionally suppressing of noise were achieved by grounded strip which is situated between electrodes. The tungsten filament serves as electron source. The filament is situated above hole in upper plate. The negative potential which is applied on upper plate create the holding electrical field $E_\perp$, which presses electrons to substrate. The density of the SE is determined by compensation of the field SE and holding field. The typically density was $n = 10^8 \, cm^{-2}$ at $E_\perp = 4 \cdot 10^2 \, V/cm$. The temperature of cell varied in range 1 – 3 K. Stabilization temperature of the cell was 1 mK and the velocity scanning of temperature was $10^{-3}$ K/c. The magnitude of temperature was definite by Speer-thermometer situated on bottom of cell.

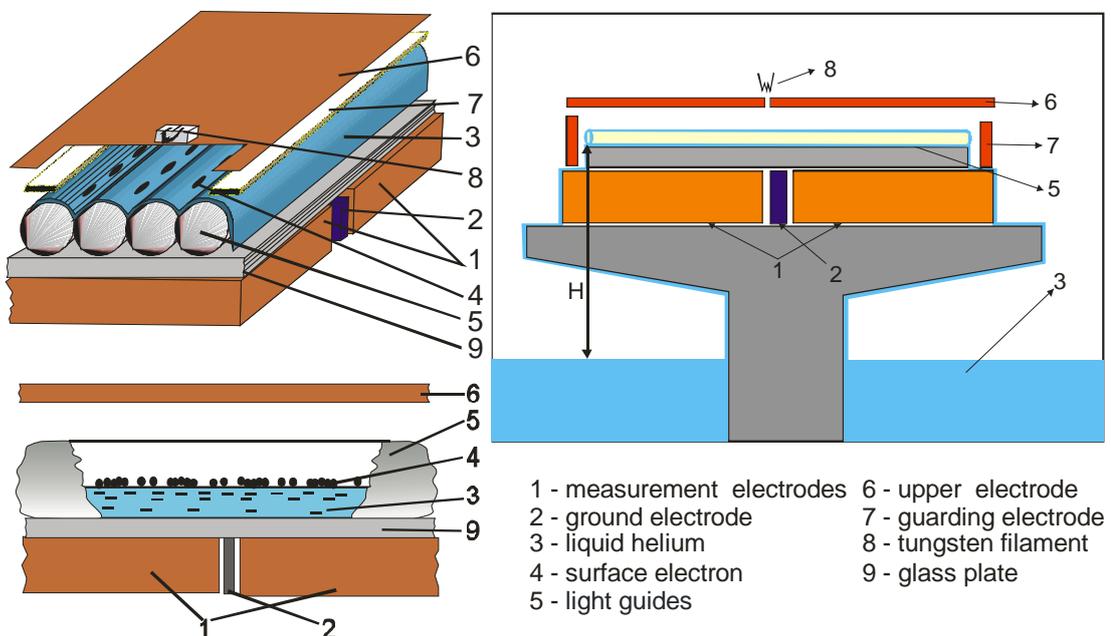

1 - measurement electrodes   6 - upper electrode
2 - ground electrode              7 - guarding electrode
3 - liquid helium                     8 - tungsten filament
4 - surface electron                9 - glass plate
5 - light guides

*Figure 1. Experimental cell.*

For study of influence of the potential variations on autolocalization quazi-one-dimensional electrons, two example of substrate is used. One of them is row of plastic light-guiders 210 µm in diameter which is placed on glass plate. The electron stripes formed in grooves between dielectric fibers over distorted surface of helium, which flowing to substrate on the height *H*.

The value of potential well for electron in groove center is definite as relation $\varphi = eE_\perp \delta - \varphi_i$ here $\delta$ is size of surface distortion, $\varphi_i$ is potential of force of the image of electron in substrate. The conducting channel width is determined as $2a = 4(eRn_l / E_\perp)^{\frac{1}{2}}$ at curvature radius of helium surface in groove $R = \alpha / \rho g H$, (where $\alpha$ and $\rho$ is the surface tension coefficient and the density of helium correspondently, $g$ is the gravity acceleration) and was near 25 μm at curvature radius 40 μm.

Other model of substrate gives possibility to form potential variation along conducting channel. In this case the light guides were covered by polymer film with two-dimensional wave-like surface (fig.2) [13]. Deflection of structure heights and period of this lattice were ~ 450 nm and 5 μm correspondently. The electron channels have been formed in region tops of fibers which covered by polymer and helium films. The magnitude $H$ was 2 mm.

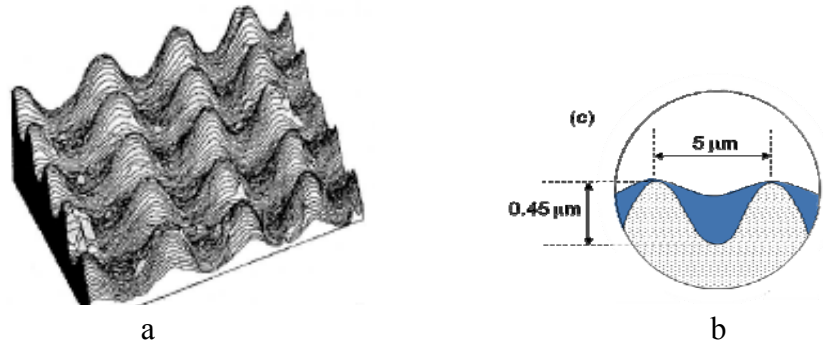

a                                                                 b

*Figure 2. a) piece of polymer film with corrugated surface; b) cross section of substrate.*

**The electrostatic calculation of dielectric cylinder in transverse electric field.** This question is considered for definition of profile of the potential well which forms electron strip over plate covering cylinder (fig. 3).

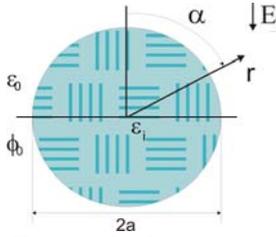

*Figure 3. Cross section of a dielectric fiber in electric field.*

Ordinary relation is $\Delta\varphi = 0$ (here $\Delta$ is Laplassian in cylinder coordinates: radial $r$, azimuthally angle $\alpha$ and longitudinal $z$). Using Fourier solution method for potential $\varphi = M(r) N(\alpha)$ (first and second derivation $\varphi$ on $z$ is zero) the equation can be presented as

$$\frac{r}{m} = \frac{d}{dr}\left(r \cdot \frac{dM}{dr}\right) + \frac{1}{N}\frac{d^2 N}{d\alpha^2} = 0 \quad (1)$$

Solution must be determined at next conditions: the potential jump on boundary is absent; normal components of electric induction inside (index $i$) and outside (index $e$) of the dielectric cylinder are equal.

$$\varepsilon_i \left(\frac{\partial \varphi_i}{\partial r}\right)_{r=a} = \varepsilon_e \left(\frac{\partial \varphi_e}{\partial r}\right)_{r=a} \quad \text{here } \varepsilon - \text{dielectric permittivity.}$$

Solution for $\varphi_e$ is

$$\varphi_e = E_\perp \left[\left(\frac{\varepsilon_i - \varepsilon_e}{\varepsilon_i + \varepsilon_e}\right) \cdot \frac{a^2}{r} - r\right] \cdot \cos\alpha + \varphi_0 \quad (2)$$

$a$ is radius of cylinder, $\varphi_o$ is initial potential on $z$- direction and $x = r \cdot \cos\alpha$ is coordinate along field.

As shown in (2) at small $\alpha$ the profile of potential well for SE over plate (like presented on Fig.3) on a fiber takes place parabolic law, which independent $H$. Estimates give the depth of potential wells for both substrates ~ $10^4$ K with approximately same width of the conducting channels.

The calibration measurements of SE mobility and density over bulk helium without substrate were performed in separate experiment.

**Results and discussion**

The temperature dependence σ(T) of Q1D - electron channels without potential variations are presented on fig 4 [10], that is typically for these condition. The other data which was obtained for wave-like substrate with variations of the potential along surface shown on fig.5. The upper curves on figs. 4, 5 reflect behavior of conductivity at increasing T and other curves - at decreasing T. In first case (fig.4) curve 1 takes place the jump down (sharp decreasing about two orders of value σ) at $T_c \sim 2.44$ K and after this was been sharp increasing at same $T_c$. These jumps are caused by appearance or disappearance of the electron autolocalization, what reflects transition of the electron transport between kinetic and hydro-dynamic regime. The disagreement of conductivity magnitudes of both curves may be caused by loss some electrons during experiment.

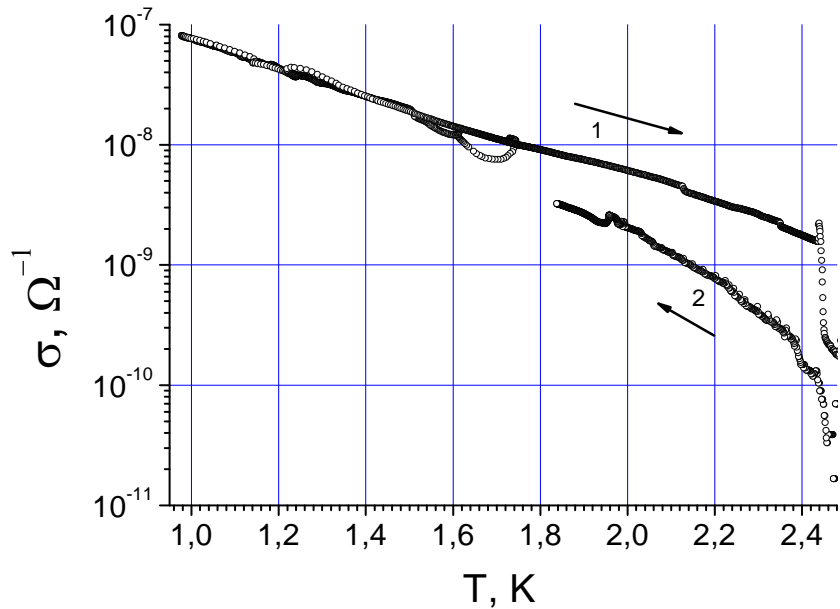

*Figure 4. Formation autolocalized state of the electrons over plastic fibers.*

On dependence σ(T) (fig.5) at $T > T_c$ take place some features when the substrate has potential variations along surface. At increasing T the magnitude of the conductivity is jump up and then decreases with T. This dependence goes to decreasing at lowering temperature too and has large jump up at same magnitude $T_c$. The lowering $T_c$ in small limit has been observed with increasing electric field. Small jump up can be seen for first substrate too.

The parabolic potential well for first substrate is [10]

$U(r) = \dfrac{m\omega_0^2 r^2}{2}$, were $\omega_0 = (eE_\perp / mR)^{1/2}$ is characteristic frequency of harmonic oscillator; $r$ is transverse coordinate. The electron wave function is

$$\psi(r) = \pi^{-1/4} r_0^{-1/2} \exp\left(-\dfrac{r^2}{2r_0^2}\right), \qquad (3)$$

where $r_0 = \left(\dfrac{\hbar}{m\omega_0}\right)^{1/2}$ is localization scale.

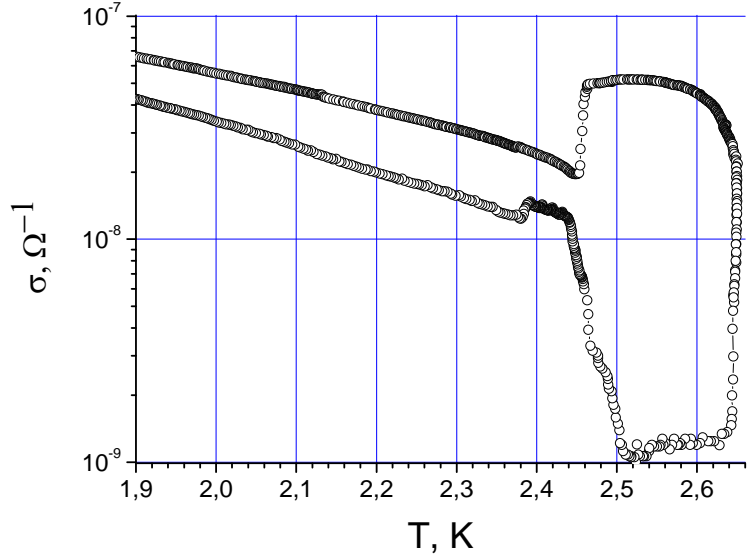

*Figure 5. Formation autolocalized state of the electrons over corrugated polymer substrate.*

The electron energy spectrum is $\varepsilon_n = (n+1/2)\hbar\omega_0$. At the experimental conditions the magnitude of this value essentially smaller than temperature and many energy levels are occupied. Thus, the electron system is not one-dimensional strongly and conducting channel is broad relatively. In this case we observe the quasi-one-dimensional electron system. The SE mobility at large relatively cross section of channel is definite by scattering SE on helium atoms in gas phase

$$\mu_g = \frac{2}{3\pi^2} \frac{e}{n_g a_0^2 \hbar \gamma}, \qquad (4)$$

here $a_0 = 0{,}62 \cdot 10^{-8}$ cm is the electron length scattering on helium atoms, $\gamma$ is opposite Bohr's radius, $n_g$ is density of gas atoms. Essential deflection of these values can be caused when appear bubbles leading to hydrodynamic moving of charge particle. The common energy state of bubble can be described by quantum mechanical energy and by energy of external forces:

$$W = W_{in} + W_{out} = \frac{\pi \hbar^2}{2 m_b r^2} + (4\pi\sigma r_b^2 + \tfrac{4}{3}\pi p r^3 + W_{sub}) \qquad (5)$$

(here $m_b$ and $r_b$ is mass and effective radius of bubble, $\sigma$ and $p$ is surface tension coefficient and external press correspondently and $W_{sub}$ is energy of interaction with substrate. For gas situation the surface tension coefficient is absent. In Q1D condition the SE autolocalization without potential variation is described in [10].

The wave-like surface of polymer film (fig.2) causes additional potential variations along channels. The helium film suspended on structure heights, such structure leads to additional compression of the electron wave function along channel. The magnitude of the helium film distortion $\delta$, in cavities is neglected and from $eE_\perp \delta$ come to 10 K. According this essential part of electrons are localized over waves where the bending energy is definite by dielectric helium film $d$, on substrate is more then 50 K at $d = 50$ nm.

Notice, that here the start formed of a electron autolocalization state which appear at $T_c = 2.48 K$. For such behavior we proposed the following explanation.

For creating of the bubble need energy scale about ~ 0.1 eV, which can be obtaine from thermostat and this energy turn back when a bubble is disappear. Even 1% electrons of passing to the autolocalization state gives a change in energy of subsystem more than $10^5$ eV, and that is essentially, because electron system thermo-insulated from thermostat. Last is caused by absent of helium superfluity, multi-layeredness of substrate, small gas density and others. This leads to essential gradient of temperature between electron subsystem and thermostat.

The heating of thermostat leads to creating bubble fraction in electron subsystem and to jump up of conductivity SE fraction. Increasing bubble fraction causes decreasing of the conductivity of electron subsystem. Process is opposite at cooling of thermostat and heating electron subsystem by disappearing bubbles. Process is self-consistent. From estimate of temperature dependence SE conductivity take a place the jump up to 1.8 K for electron subsystem at heating thermostat and at appear bubble fraction. In opposite that disappear bubbles can be lead to heating of the electron subsystem up to 3 K at disappear bubbles. These features absent practically for experiments with 1-type of substrate, where electron subsystem is connected with thermostat stronger.

## Conclusion

Auto localization of quasi-one-dimensional electrons in a gas phase over helium with or without variations of potential is investigated in work. The substrate represented a number of dielectric fiber with or without a polymeric film on surface. The carried out electrostatic analysis of a substrate of the second type in a constant field. Showed possibility of formation of quasi-one-dimensional electrons on a film over light guides. Wave-like surface of polymer film induces of potential variations in conducting channels. Appearance and disappearance of autolocalization of SE leads to jump down or jump up of mobility for both substrates. Takes place features in the presence of polymer film. At increasing T the mobility jumps up and then smoothly decreases with T. This dependence goes to decreasing at lowering temperature and then has large jump up at same magnitude $T_c$. Features is explained by authors as dynamic process caused by changing subsystem energy at appear or disappear electron auto-localization and by thermo-insulation electron system from thermostat.